# Effect of Interfacial Dipole on Heterogeneous Ice Nucleation

Hao Lu, Quanming Xu, Jianyang Wu, Rongdun Hong[*], Zhisen Zhang[*]

Department of Physics, Research Institute for Biomimetics and Soft Matter, Fujian Provincial Key Laboratory for Soft Functional Materials Research, Jiujiang Research Institute, Xiamen University, Xiamen, Fujian 361005, China

**ABSTRACT:** In this letter, we performed molecular dynamics simulations of ice nucleation on a rigid surface model of cubic zinc blende structure with different surface dipole strength and orientation. It follows that, despite the excellent lattice match between cubic ice and substrates, the ice nucleation happened only when the interfacial water molecules (IWs) have the same or similar orientations as that of the water molecules in cubic ice. The free energy landscapes revealed that, for substrates with improper dipole strength/orientation, large free energy barriers arose to prevent the dipole of IWs rotating to the right orientation to trigger ice formation. Our results suggest that the traditional concept of lattice match, the similarity of lattice length between a substrate and the new-formed crystalline, should be extended to a broader match include the similarity between the molecular orientations of the interfacial component and the component in the specific new-formed crystalline face.

**TOC Graphic**

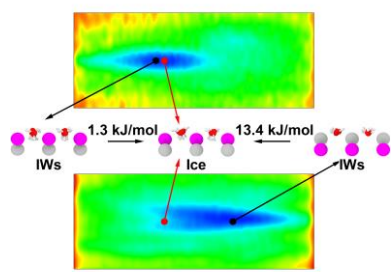

## INTRODUCTION

Ice nucleation is important in many ways, such as climate, microbiology, and atmospheric science.[1-5] Due to the lower free energy barrier of heterogeneous nucleation, in practice, the freezing process of water is usually controlled by heterogeneous nucleation caused by an external medium. This external medium is called ice nucleating particle (INP), and many substrates have been proven to be effective heterogeneous INPs, such as dust,[6] cholesterol,[7-8] silver iodide,[1-2, 9-11] kaolinite,[12-15] etc.

According to the classical nucleation theory (CNT), lattice match between INPs and ice is considered to be the main reason for promoting ice nucleation in traditional views.[1, 16-19] Kaolinite and silver iodide are treated as effective INPs in ice nucleation for the Al-surface of kaolinite has a potential crystallographic match with hexagonal ice; $\gamma$-AgI has a potential crystallographic match with cubic ice.[14, 17] However, recent researches shows that, only with lattice match, it is insufficient to predict the ability of INPs in ice nucleation.[1-2, 20-24] Zielke and Bertram[17] pointed that ice nucleation occurs only when silver ions are exposed on the AgI surface, rather than iodide ions exposed on the surface of AgI. Experiments also reported differences in the icing temperatures between $Ag^+$ enriched side and $I^-$ enriched side.[25] Furthermore, no ice nucleation events observed in any base plane of barium fluoride ($BaF_2$) even there is a good crystallographic match between $BaF_2$ and ice crystals.[26-27] Hydrophobicity,[22, 28] electric field[16, 29] obtained additional consideration beyond lattice match in the process of ice nucleation. Shao et al[26] reported that the combination of the oxygen lattice order and the hydrogen disorder of IWs on substrates can effectively facilitate the freezing of top water.

[*] Corresponding Emails: zhangzs@xmu.edu.cn, rdhong@xmu.edu.cn

In this work, a substrate model with cubic zinc blende structure and different surface dipole strength and orientation, which possessed a fine crystallographic match with ice, was adopted to investigate the effect of interfacial dipole on heterogeneous ice nucleation. Free energy landscapes of dipole orientation of IWs were calculated to describe the ability of different substrates in promoting ice nucleation. In the subsequent simulation by restricting the behavior of IWs, we confirm that the orientantion of IWs is the key to unlock the secret of ice nucleation.

## SIMULATION DETAILS

Practically, many materials take the cubic zinc blende structure, e.g. cadmium sulfide (CdS)[30], $\gamma$-AgI[31] (which is well known for its ability in promoting ice formation[17]). Thus, the lattice parameters of $\gamma$-AgI, namely $a = b = c = 6.50$ Å, $\alpha = \beta = \gamma = 90°$ (with a lattice mismatch of 2.36% compared to cubic ice, which are $a = b = c = 6.35$ Å, $\alpha = \beta = \gamma = 90°$),[32-33] were taken as a substrate model to investigate heterogeneous ice nucleation. The simulation cell consisted of two mirrored substrate slabs to eliminate the long-range electric field produced by lattice truncation.[17, 34-35] The slabs were located 4.80 nm for placing 4032 water molecules, and a 3.00 nm vacuum area located outside the two slabs. The box's dimension was $5.20 \times 5.20 \times 13.000$ nm$^3$ (Supporting Information, Figure S1). Periodic boundary conditions (PBCs) were applied in all directions.

All MD simulations were carried out in the NVT ensemble employing the GROMACS 4.6.7 package.[36] The TIP4P/Ice[37] water model (melting point 270 ± 3 K) was applied for the water molecules. LINCS algorithm was employed to restrict hydrogen bond lengths.[38] A time step of 2 fs was employed. The particle-mesh Ewald[39] method was used to calculate the long-range Coulomb interactions, with a cutoff of 1.3 nm for the separation of the direct and reciprocal space summation. The cutoff distances for vdW interactions were set to 1.3 nm. Nose-Hoover thermostat was used to maintain the temperature.[40-41]

**Ice Formation on Different Substrates.** The temperature in the simulations was performed at 300 K for 1 ns and then decreased to 250 K, namely 20 K below the TIP4P/Ice water model melting point, for the rest of the simulation time. Surface dipole strength was adjusted by the partial charges of cations/anions, ranging from 0.0e to 1.0e (refer to Table 1 for more details). For each system, $10 \times 200$ ns MD trajectories were carried out.

**Density Distribution Profiles of IWs.** Density distribution profiles of hydrogen and oxygen atoms along $z$-axis (perpendicular to the surface of substrate) were calculated to reveal the behavior of IWs (at 300 K). The simulation box was divided into 1000 slices to calculate the density of hydrogen and oxygen atoms in each slice by *g_density* in GROMACS.

**2-Dimension (2D) Free Energy Landscape of IWs Dipole Orientation.** The 2D free energy landscapes as a function of the angle of $z$-axis/IWs dipole orientation and the position of $x$-axis were calculated by metadynamics method[42-43] (GROMACS 4.6.7 with PLUMED 2.2.3[44]). Walls were applied to enhance the sampling (Figure S2). The force constant for the walls KAPPA were set to 1500 kJ mol$^{-1}$ nm$^{-2}$. In metadynamics simulation, the widths of the Gaussian hills SIGMA were set to 0.04. The heights of the Gaussian hills HEIGHT were set to 2.0 kJ/mol. The output frequency for hill addition was set to 500 steps.

## RESULTS AND DISCUSSION

Figure 1 shows the snapshots of simulations at different times on cation-exposed (+0.4e, Figure 1A,

1B) and anion-exposed surface (-0.4e, Figure 1C, 1D). The freezing of water molecules on the cation-exposed surface only takes a short time (about 15 ns) (refer to ESI Movie 1). However, the water molecules on the anion-exposed surface maintain the liquid state even after 200 ns (Figure 1D) (refer to Figure S3 and Movie 2). The substrates, with almost perfect lattice match with ice, exhibit completely different promotion effects on heterogeneous ice nucleation. Therefore, the surface dipole, as the only difference between the two substrates, received additional consideration.

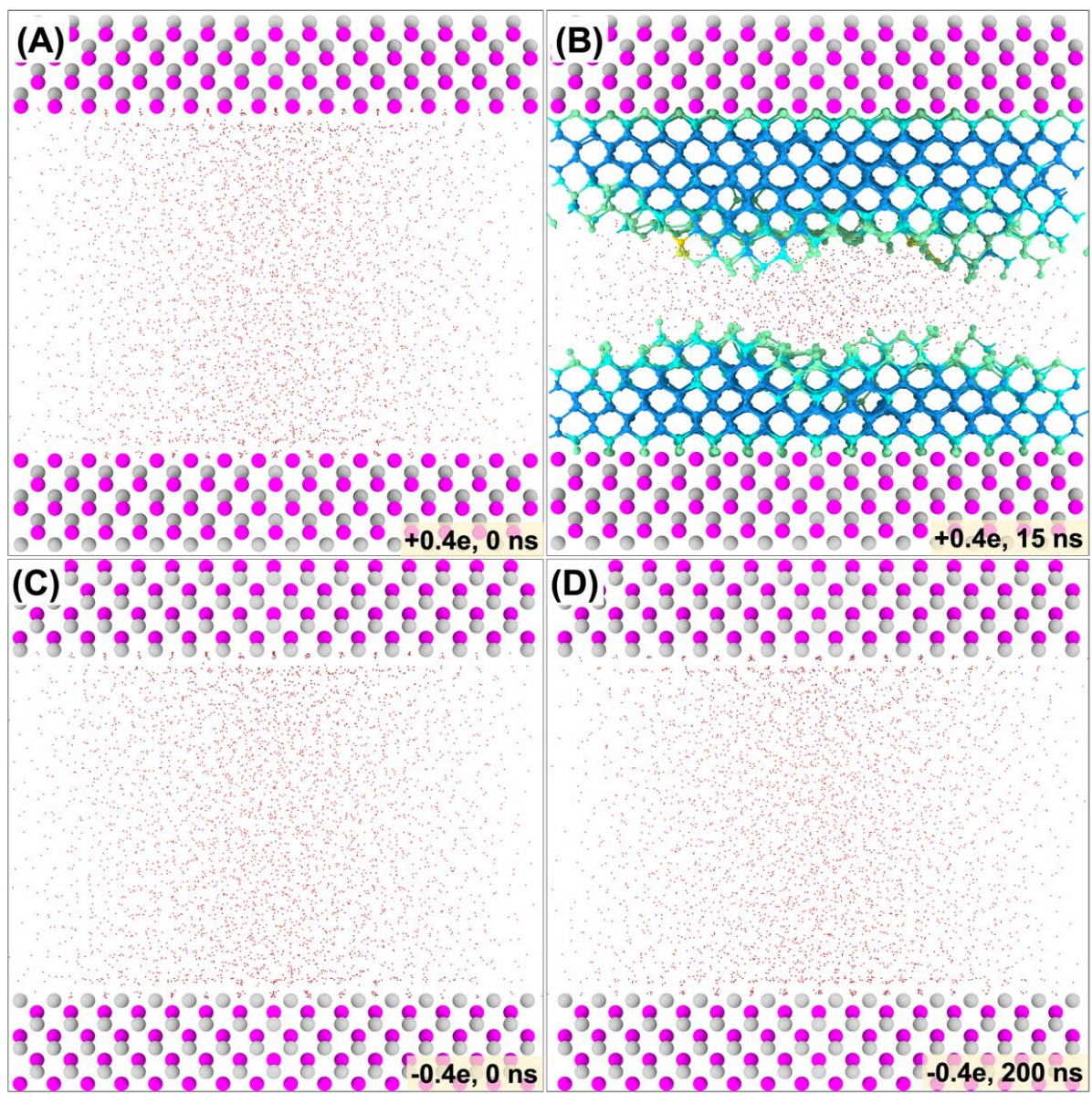

Figure 1. Snapshots of ice formation on cation-/anion-exposed surfaces. Panels A and B are cation-exposed as the outermost layers at 0 ns and 15 ns. Panels C and D are anion-exposed as the outermost layers at 0 ns and 200 ns. For clarity, only oxygen atoms are shown in TIP4P/Ice water model. Cations, anions, and oxygen atoms in liquid water are colored by purple, grey, and red, respectively. Oxygen atoms in ice crystallites are represented by the blue and cyan CPK model. The partial charges of cations/anions are ± 0.4e.

To explore how the interfacial dipole affects ice nucleation, as shown in Table 1, 10 × 200 ns MD trajectories were performed for all the systems with different anion/cation partial charges. The partial charges of cations/anions are varied from 0.0e to 1.0e with an interval of 0.1e. The observed freezing events are marked by " √ ". The cation-exposed surface with the partial charges of cations range from +0.2e to +0.6e can nucleate ice in all the 10 MD trajectories, which is consistent with the previous reports.[17] However, there were no ice nucleation events observed when the partial charges of exposed cations/anions beyond the range of +0.2e to +0.6e. From these results, it can be concluded that the interfacial electrostatics can significantly alter the promotion effect of substrate on heterogeneous ice nucleation.

Table 1. Freezing condition at different partial charges on cation-/anion-exposed surface. " √ " represents ice nucleation events observed within 200 ns MD simulation. "-" represents no ice nucleation events observed within 200 ns MD simulation.

| Partial Charges | 0.1e | 0.2e | 0.3e | 0.4e | 0.5e | 0.6e | 0.7e | 0.8e | 0.9e | 1.0e |
|---|---|---|---|---|---|---|---|---|---|---|
| Cation | - | √ | √ | √ | √ | √ | - | - | - | - |
| Anion | - | - | - | - | - | - | - | - | - | - |

As mentioned above, the interfacial electrostatics can severely alter the heterogeneous ice nucleation. Moreover, it has been reported that the behavior of IWs played a key role on heterogeneous ice nucleation.[26, 45-49] Given the strong dipole of water molecules, it's reasonable to believe that, in our MD simulations, the behavior of IWs was altered by the interfacial electrostatics on different substrates, and hence affect heterogeneous ice nucleation. To explore the mechanism of how interfacial electrostatics affects heterogeneous ice nucleation, the density distribution profiles (along $z$-axis) of hydrogen and oxygen atoms in IWs were extracted (shown in Figure 2). The detailed configurational snapshots of the IWs are shown in Figure 2D-F.

As shown in Figure 2A and 2C, the density distribution profiles of hydrogen/oxygen atoms in IWs are significantly different from those in ice. For the IWs configurations in Figure 2D, 2F, the dipole orientations of IWs are parallel to the $z$-axis. For these two systems, there was no ice nucleation event observed even in 10 × 200 ns trajectories. While in Figure 2B, the density distribution profiles of hydrogen/oxygen atoms in IWs are almost the same as those in the substrate-induced ice (formed in the system of +0.4e system at 250 K). For the IWs configurations in Figure 2E, one of the H-O bond in IWs is parallel to the $z$-axis, making an angle of 1.19 rad between the $z$-axis and dipole orientation of IWs, which is almost the same as the water molecules in the substrate-induced ice. Within 15 ns, the number of water molecules freezing to ice reached the maximum value. Thus, the different behaviors of IWs in Figure 2D-F resulted in totally different ice formation processes. The heterogeneous ice nucleation process can be roughly broken down into two steps: i) the formation of ice-like structure (Figure 2E) on the substrate; ii) the accumulation of other water molecules around the ice-like structure to form ice.[16] In the cubic zinc blende structure, the distances between two neighboring cations/anions (in $x$-/$y$-axis directions) are 6.50 Å, which are almost the same as the O-O distances between two neighboring water molecules in (001) face of ice, namely 6.35 Å (refer to Figure 2G). While, to nucleate ice on the substrate surface, besides the lattice match between substrate and ice, there is another requirement: the orientation of IWs should be also similar to that in (001) face of ice, which is severely dependent on the interfacial electrostatics. Thus, although an almost perfect lattice match was achieved in all the simulation systems here, the ice formation events were observed only in a few systems.

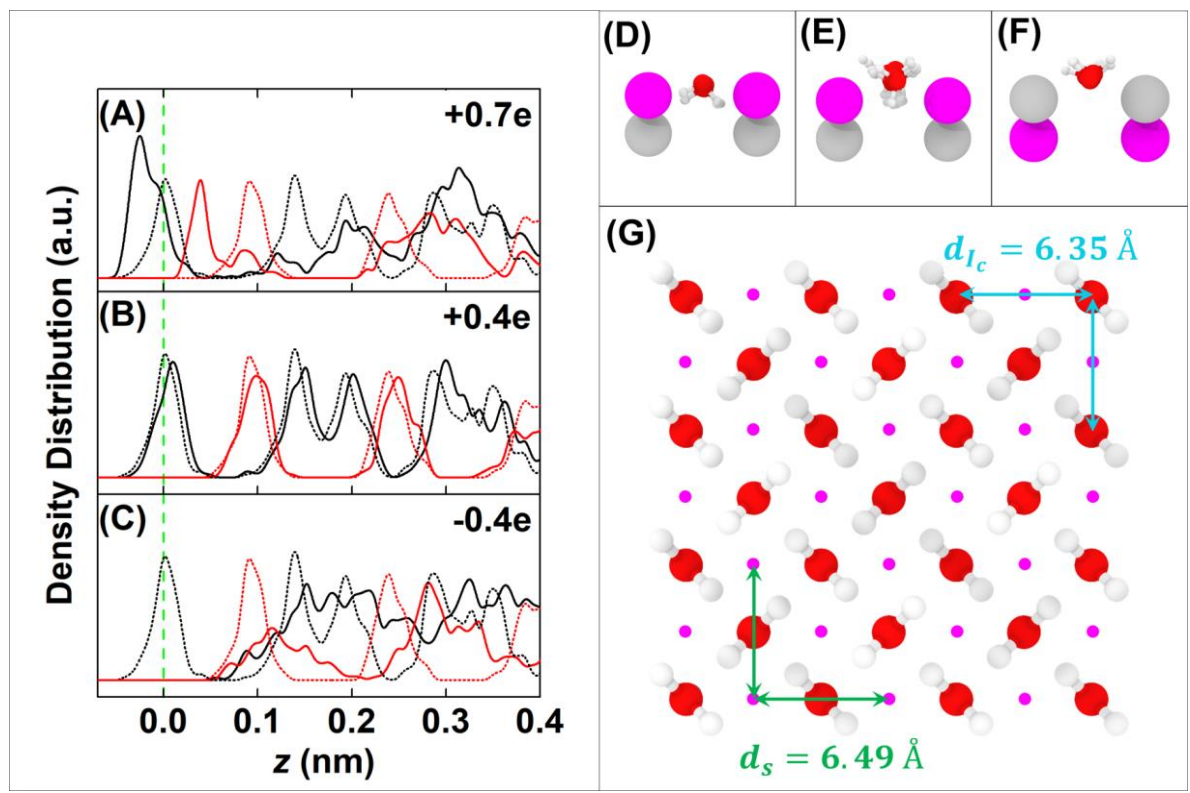

Figure 2: Density distribution profiles of hydrogen/oxygen atoms in water molecules on the surface of substrates along $z$-axis. The partial charges of the exposed cations/anions on the surface in panels A-C are +0.7e, +0.4e, and -0.4e, respectively. The profiles of hydrogen atoms and oxygen atoms are black and red. The dot lines of the density distribution of hydrogen/oxygen atoms in substrate-induced ice (formed in the system of +0.4e system at 250 K) in panels A-C are black and red, respectively. The vertical green dotted line coincides with the position of the outermost cations/anions on the surface. Panels D-F are configuration snapshots of IWs corresponding to panels A-C. Panel G is the configurations of exposed cations/anions on the surface and water molecules in (001) face of Ic. In panel E-G, cations, anions, hydrogen atoms, and oxygen atoms are colored by purple, grey, white, and red, respectively.

To quantitatively investigate the impact of IWs behavior on heterogeneous ice nucleation, the 2D free energy landscapes of IWs' dipole orientation were calculated (Figure 3). As shown in Figure 3, the 2D free energy landscape of IWs' dipole orientation exhibits a periodic structure along the $x$-axis, with the maxima corresponding to the surface cations/anions and the minima corresponding to the adsorption sites (see Figure 2D-F). For the 2D free energy landscapes of system +0.4e (Figure 3A), the 2D free energy minima locate at the angle (between $z$-axis and dipole orientation of water molecules) of 1.19 rad, which is almost the same as that of IWs in substrate-induced ice, 1.15 rad. The free energy barrier is about 1.3 kJ/mol for the adsorbed IWs rotating to an ice-like structure ((001) face in Ic). While, for the system of -0.4e (Figure 3B), the 2D free energy minima locate at the angle of 2.14 rad, which is far from 1.19 rad. Moreover, the free energy barrier is found to be 13.4 kJ/mol for the adsorbed IWs changing into an ice-like structure, making the ice nucleation process on the substrate much harder, which is consistent with the results of ice formation simulations that ice nucleation events were not observed in 10 × 200 ns trajectories for the -0.4e system. These results confirmed that the interfacial electrostatics of different substrates can severely alter the behavior of IWs, and hence affect heterogeneous ice nucleation.

It should be pointed out that although the substrates can not nucleate ice in 200 ns (15 out of 20 systems in Table I), it does not mean that they will never nucleate ice. For these substrates, ice could nucleate under a much larger supercooling or in a much longer time. For instance, the free energy barrier of -0.4e system is 12.1 kJ/mol larger than that of +0.4 system (about 5.82 $k_B$T at 250 K), which will result in an about 337 ($e^{5.82}$) times smaller heterogeneous ice nucleation rate. Besides, in a previous experimental study,[25] the authors found that the icing temperature of $Ag^+$ side of poled wurtzite AgI is -2.7 ± 0.4 ℃; while the icing temperature is -4.2 ± 0.4 ℃ for that $I^-$ side of poled wurtzite AgI.

These results are consistent with the classical nucleation theory (CNT). According to CNT, the heterogeneous premotion effect of a substrate depends on the interfacial free energy differences between the fluid phase and the crystal phase on the substrate.[5] For a solid crystalline, the interfacial free energy between the crystalline phase and substrate $\gamma_{cs}$ is highly dependent on the misorientation angle $\varphi^5$. Thus, to predict the heterogeneous nucleation promotion effect of substrate, the concept of lattice match should be extended to structure match, including the lattice match and molecular orientation match between the interfacial component of substrate and the new-formed crystalline face on the substrate.

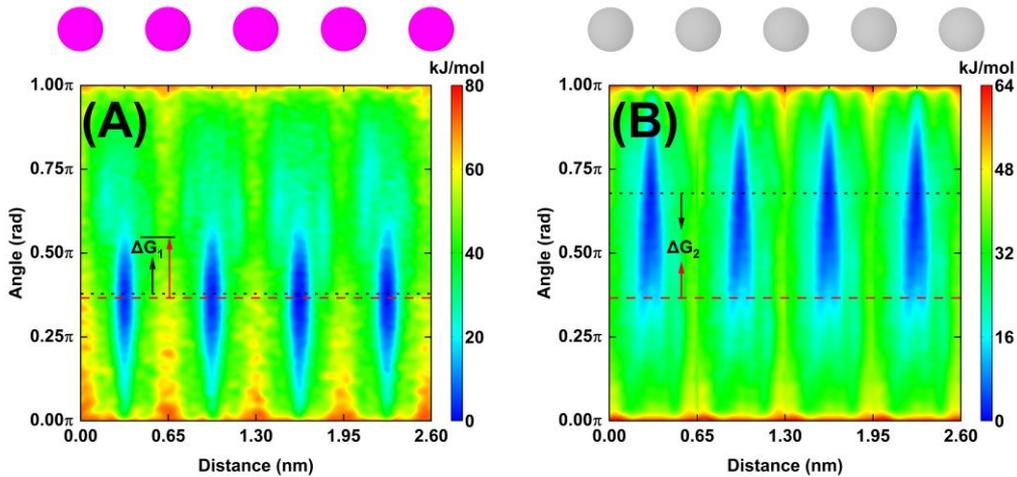

Figure 3: Free energy landscapes of IWs' dipole orientation as a function of Angle (between dipole orientation of IWs and $z$-axis) and Distance (between coordinate of IWs and center of simulation box in $x$-axis). The partial charges of cations/anions on the surfaces in panel A and panle B and are +0.4e and -0.4e. Black dotted lines refer to the angle corresponding to the lowest point of the free energy landscape. Red dashed lines represent the angle between the dipole orientation of the IWs in substrate-induced ice and the $z$-axis. Δ$G$ refers to the difference of free energy corresponding to these two states, Δ$G_1$=1.3 kJ/mol, Δ$G_2$=13.4 kJ/mol. Cations and anions are colored by purple and grey.

Based on the results of ice formation simulations, IWs adsorbing configurations and 2D free energy landscapes, we emphasized that, with a suitable interfacial dipole strength and orientation, the substrate can guide water molecules at the interface to adjust their adsorption configurations to form an ice-like structure, and then induced the formation of bulk ice. To further verify this point, the equilibrium adsorption configuration of IWs (before freezing) in +0.4e system was extracted, fixed and applied to the systems of -0.4e, -0.2e, 0e, +0.2e, +0.4e, and +0.7e.

With non-fixed IWs on the substrates, for systems of -0.4e, -0.2e, 0e, and +0.7e, there was no ice formation observed in 10 × 200 ns MD simulation trajectories (Table 1 and Figure 4A-C, F). While, with fixed IWs on the substrates, ice formation was observed in all 10 × 200 ns MD simulation trajectories for all the systems motioned above, regardless of the partial charges of the exposed cations/anions. The snapshots of ice formation trajectories at 15 ns in all systems are shown in Figure 4(a)-(f). For the systems of +0.2e and +0.4e (Movie 3), there are more frozen water molecules than that in systems of -0.4e (Movie 4), -0.2e, 0e, and +0.7e, illustrating that, with the fixed IWs, the partial charges of the exposed cations/anions can only slightly alter the speed of ice formation, instead of preventing the formation of ice. Thus, the interfacial electrostatics indirectly affects the heterogeneous nucleation of ice by regulating the adsorption behavior of IWs.

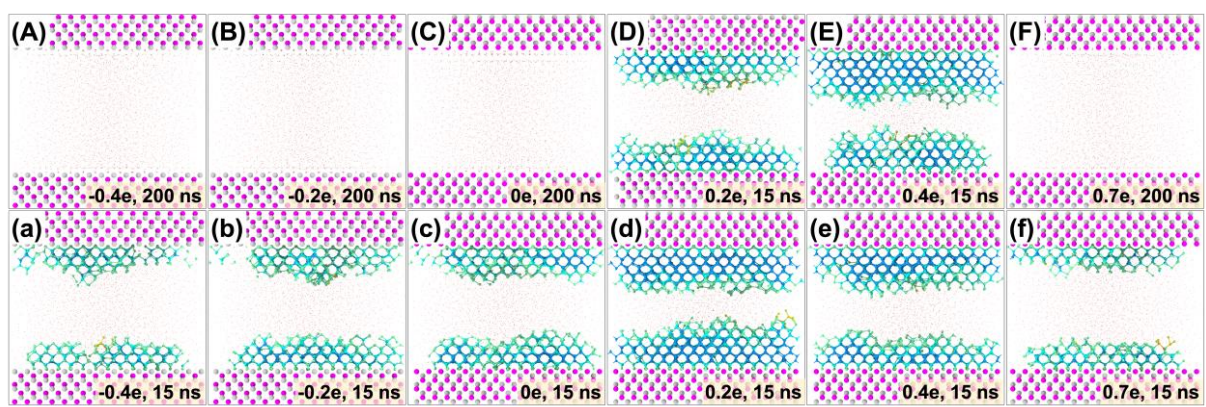

Figure 4: Ice formation events in several systems with different substrate dipole strength and orientation. Interfacial water molecules were not fixed in Panel A-F. While, interfacial water molecules were fixed in Panel (a)-(f). For clarity, only oxygen atoms are visible in TIP4P/Ice water molecules. The partial charges of cations/anions and simulation time are shown in the bottom of each panel. Cations, anions, and oxygen atoms are colored by purple, grey, and red, respectively.

## CONCLUSIONS

We performed MD simulations of ice nucleation on a rigid surface model of cubic zinc blende structure with different surface dipole strength and orientation. It follows that the substrate dipole (strength and orientation) guides the behavior of water molecules at the interface, thereby affecting ice nucleation. We verified that the traditional concept of lattice match is insufficient to predict the ice nucleation promotion effect of substrate even when the substrate possesses an almost perfect lattice match with ice. The concept of lattice match should not only contain the matching of the substrate and new-formed crystalline, but also should be extended to include the matching between the orientation of interfacial component and new-formed crystalline face.

## ASSOCIATED CONTENT

**Supporting Information**

The Supporting Information is available free of charge at http://pubs.acs.org.

System setup (Figure S1); Walls setting of Metadynamics (Figure S2); Ice Formation on Different Substrates (Figure S3). (PDF)

Dynamic diagram of simulation trajectory without fixing the Interfacial water molecules (The partial charges of exposed cations/anions on the surface are +0.4e). (Movie_1_+0.4e) (AVI)

Dynamic diagram of simulation trajectory without fixing the Interfacial water molecules (The partial charges of exposed cations/anions on the surface are -0.4e). (Movie_2_-0.4e) (AVI)

Dynamic diagram of simulation trajectory after fixing the Interfacial water molecules (The partial charges of exposed cations/anions on the surface are +0.4e). (Movie_3_+0.4e-fixed-IWs) (AVI)

Dynamic diagram of simulation trajectory after fixing the Interfacial water molecules (The partial charges of exposed cations/anions on the surface are -0.4e). (Movie_4_-0.4e-fixed-IWs) (AVI)

## AUTHOR INFORMATION

### Corresponding Authors

* E-mail address: zhangzs@xmu.edu.cn
*E-mail address: rdhong@xmu.edu.cn

### Notes

The authors declare no competing financial interest.

## ACKNOWLEDGEMENTS

This work is financially supported by the National Natural Science Foundation of China (Grant Nos. 11904300, 11772278 and 11502221), the Jiangxi Provincial Outstanding Young Talents Program (Grant No. 20192BCBL23029), the Fundamental Research Funds for the Central Universities (Xiamen University: Grant Nos. 20720180014, 20720180018 and 20720180066). Y. Yu and Z. Xu from Information and Network Center of Xiamen University for the help with the high-performance computer clusters.